%% file: main.tex
\documentclass[sigconf]{acmart}

\renewcommand\footnotetextcopyrightpermission[1]{}
\input{utils/packages}      
\input{utils/def}           

\urldef{\mailsa}\path|nravi@hpe.com|

\begin{document}

\title{Exploring Fully Offloaded GPU Stream-Aware Message Passing}
\author{Naveen Namashivayam}
\email{naveen.ravi@hpe.com}
\affiliation{%
    \institution{Hewlett Packard Enterprise}
    \country{USA}
}
\author{Krishna Kandalla}
\email{krishnachaitanya.kandalla@hpe.com}
\affiliation{%
    \institution{Hewlett Packard Enterprise}
    \country{USA}
}
\author{James B White III}
\email{trey.white@hpe.com}
\affiliation{%
    \institution{Hewlett Packard Enterprise}
    \country{USA}
}
\author{Larry Kaplan}
\email{larry.kaplan@hpe.com}
\affiliation{%
    \institution{Hewlett Packard Enterprise}
    \country{USA}
}
\author{Mark Pagel}
\email{mark.pagel@hpe.com}
\affiliation{%
    \institution{Hewlett Packard Enterprise}
    \country{USA}
}

\input{src/content/abstract}

\maketitle


\input{src/content/introduction}

\input{src/content/solution}
\input{src/content/background}

\input{src/content/proposal}
\input{src/content/implementation}
\input{src/content/performance}
\input{src/content/related}
\input{src/content/conclusion}

\balance
\bibliography{main}

\end{document}

%% file: utils/packages.tex
\usepackage{verbatim}
\usepackage{graphicx}
\usepackage{url}
\usepackage{epstopdf}
\usepackage{float}
\usepackage{graphicx}
\usepackage{wrapfig}
\usepackage{comment}
\usepackage{csquotes}
\usepackage{url}
\usepackage{xcolor,colortbl}
\usepackage{color}
\usepackage{fancyvrb}
\usepackage{epsfig}\restylefloat{table}
\usepackage{setspace}
\usepackage{multirow}
\usepackage{threeparttable}
\usepackage{mathtools}
\usepackage[ruled, linesnumbered, vlined]{algorithm2e}
\usepackage{algpseudocode}\algrenewcommand\textproc{}
\usepackage{listings}
\usepackage{listings}
\usepackage[misc]{ifsym}
\usepackage[nolist]{acronym}
\usepackage{tikz}
\usepackage{flushend}
\usepackage{framed}
\usepackage{subcaption}

%% file: utils/def.tex
\algdef{SE}[DOWHILE]{Do}{doWhile}{\algorithmicdo}[1]{\algorithmicwhile\ #1}%

\usetikzlibrary{calc}
\makeatletter
\newlength{\mylength}
\xdef\CircleFactor{1.1}
\newsavebox{\mybox}
\newcommand*\circled[2][draw=blue]{\savebox\mybox{\vbox{\vphantom{WL1/}#1}}\setlength\mylength{\dimexpr\CircleFactor\dimexpr\ht\mybox+\dp\mybox\relax\relax}\tikzset{mystyle/.style={circle,#1,minimum height={\mylength}}}
\tikz[baseline=(char.base)]
\node[mystyle] (char) {#2};}
\makeatletter

%% file: src/content/abstract.tex

\begin{abstract}
Modern heterogeneous supercomputing systems are comprised of CPUs, GPUs, and
high-speed network interconnects. Communication libraries supporting efficient
data transfers involving memory buffers from the GPU memory typically require
the CPU to orchestrate the data transfer operations. A new offload-friendly
communication strategy, \textit{stream-triggered (ST) communication}, was explored
to allow offloading the synchronization and data movement operations from the
CPU to the GPU. A Message Passing Interface (MPI) one-sided active target
synchronization based implementation was used as an exemplar to illustrate the
proposed strategy. A latency-sensitive nearest-neighbor microbenchmark was used
to explore the various performance aspects of the implementation. The offloaded
implementation shows significant on-node performance advantages over standard
MPI active RMA (36\%) and point-to-point (61\%) communication. The current
multi-node improvement is less (23\% faster than standard active RMA but
11\% slower than point-to-point), but plans are in progress to purse further
improvements.
\end{abstract}

%% file: src/content/introduction.tex
\section{Introduction}\label{sec:introduction}

To accommodate modern heterogeneous supercomputing systems comprised of CPUs and
GPUs~\cite{frontier,Larrea2019ScalingTS,sierra,CORAL}, current-generation
scientific applications and systems-software stacks use \textit{GPU-aware}
communication libraries~\cite{GPU-aware-MVAPICH}. GPU-aware libraries support
performing inter-process communication operations involving GPU buffers 
without having to stage them through a CPU-attached memory. Remote
direct memory access (RDMA)~\cite{GPU-aware-MVAPICH-inter-node,GPU-aware-MVAPICH-inter-node-2}
and vendor-specific peer-to-peer~\cite{CUDA-IPC,GPU-aware-intra-node} data transfer
mechanisms are used to implement GPU-aware inter-node and intra-node data
movement operations.

\begin{figure}[!ht]
  \includegraphics[width=\linewidth]{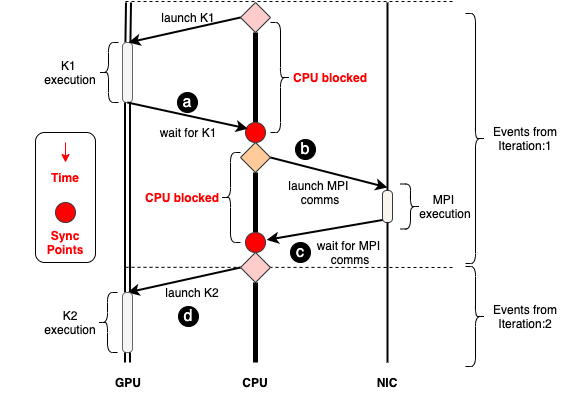}
  \caption{Events on a typical GPU-aware application.}
  \label{fig:normal-execution}
\end{figure}

Even with GPU-awareness in the communication stack, CPU threads are still
required to orchestrate data-moving communication and inter-process
synchronization operations. Figure~\ref{fig:normal-execution}
demonstrates this issue using a typical GPU-aware parallel application based on
the Message Passing Interface (MPI)~\cite{mpi} as an example for the
inter-process communication.

An MPI process running on the CPU launches a device kernel (K1) and waits for
its execution to complete.
\begin{tiny}\circled[text=white,fill=black,draw=black]{a}\end{tiny}
It synchronizes with the local GPU to ensure completion of compute
kernel (K1). Next, the CPU
\begin{tiny}\circled[text=white,fill=black,draw=black]{b}\end{tiny} launches and
\begin{tiny}\circled[text=white,fill=black,draw=black]{c}\end{tiny} waits for
completion of
the inter-process communication operation involving application GPU buffers previously
updated by the kernel (K1). To reuse GPU buffers on subsequent compute
kernel (K2),
\begin{tiny}\circled[text=white,fill=black,draw=black]{d}\end{tiny}
K2 is launched only after the inter-process communication operations have
completed.

This results in all communication and synchronization operations occurring at
GPU kernel boundaries, and creates potentially expensive synchronization points
that require the CPU to synchronize with the GPU and
Network Interface Controller (NIC) devices.

A
communication strategy called
\textit{stream-triggered (ST) communication} is 
explored in this
work. 
Along with managing compute kernels defined by the application, the proposed
ST solution allows the GPU to coordinate with the NIC to manage MPI
communication operations. This design goal avoids using the CPU cycles to
orchestrate and drive the MPI communication operations. This eliminates any
potential synchronization points between an application process and its GPU
device.

\subsection{Contributions of This Work}
\label{subsec:contrib}

The following are the major contributions of this work.

\begin{enumerate}
    \item Propose a new communication strategy to fully offload the communication
    control paths from the CPU to the GPU;
    \item Use an MPI one-sided active-target-synchronization~\cite{Hoefler-rma} based
    solution to demonstrate the proposed strategy;
    \item Implement the proposed strategy on a modern NIC, \textbf{HPE Slingshot
    NIC}~\cite{slingshot2,slingshot1,slingshot}, exploiting the supported
    \textit{triggered operations} feature~\cite{quadrics-tops,tops-portal};
    \item Discuss various implementation design options to fully offload the
    communication control path to the GPU; and
    \item Analyze the performance of the proposed ST strategy using a
    latency-sensitive microbenchmark kernel called \textit{Faces}, inspired by
    the nearest-neighbor communication pattern from the
    CORAL-2~\cite{CORAL,coral2-bm} \textit{Nekbone benchmark}~\cite{nekbone-bm}.
\end{enumerate}

%% file: src/content/solution.tex

\section{Proposed Communication Scheme}\label{sec:solution}
Communication operations in GPU-aware applications are typically comprised
of \textit{data paths} and \textit{control paths}. Data paths refer to those
operations that involve moving data between the CPU-attached and the GPU-attached memory
regions. These data movement operations can occur within the same compute node
or between different compute nodes across a high-speed network. Control paths
correspond to coordination operations that occur between (1) the application
process running on the CPU,
(2) application compute kernels running on the GPU, and (3) the network interface controller (NIC).
The proposed ST strategy enables a GPU-aware application to offload the control
paths to underlying implementation and hardware components.

\subsection{GPU Streams}\label{subsec:streams}
A GPU stream~\cite{cuda-stream} is a queue of device operations. GPU compute
kernel concurrency is achieved by creating multiple concurrent streams.
Operations issued on a stream typically run asynchronously with respect
to the CPU and operations enqueued in other GPU streams. Operations in a given
stream are guaranteed to be executed in FIFO order. In this work, the GPU
component that provides these execution guarantees to schedule and control the
execution of the enqueued operation is referred to as the \textit{GPU Stream
Execution Controller} (GPU SEC). Depending on the GPU vendor, GPU SEC can be a
software, hardware, or kernel component associated with the GPU.

\subsection{Stream-Triggered Communication}\label{subsec:stcomms}
A parallel application using the ST strategy continues to manage compute kernels
on the GPU via existing mechanisms. In addition, the new strategy allows an
application process running on the CPU to define a set of ST communication
operations. These communication operations can be scheduled for execution at a
later point in time. More importantly, in addition to offering a deferred
execution model, ST enables the GPU to be closely involved in the control
paths of the MPI communication operations. 

\begin{figure}[ht]
  \includegraphics[width=\linewidth]{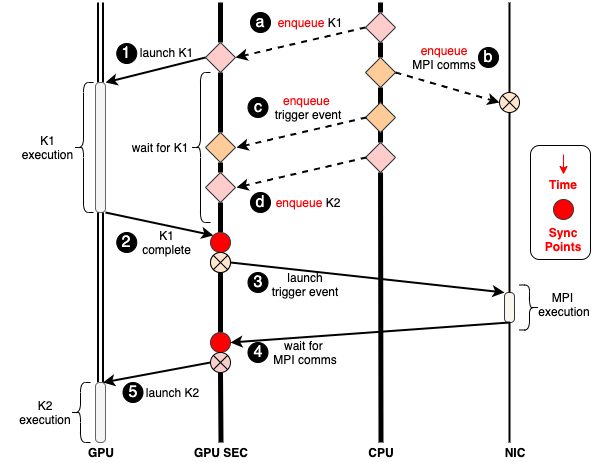}
  \caption{Events on a GPU-aware application using ST.}
  \label{fig:st-execution}
\end{figure}

Figure~\ref{fig:st-execution} illustrates a sequence of events involved in a
parallel application using MPI-based ST inter-process communication and
synchronization operations. An application process running on the CPU enqueues
\begin{tiny}\circled[text=white,fill=black,draw=black]{a}\end{tiny} GPU
kernel K1 to the GPU stream,
\begin{tiny}\circled[text=white,fill=black,draw=black]{b}\end{tiny} triggered
ST-based MPI operations to the NIC,
\begin{tiny}\circled[text=white,fill=black,draw=black]{c}\end{tiny} the
corresponding trigger event to the GPU stream, and
\begin{tiny}\circled[text=white,fill=black,draw=black]{d}\end{tiny} GPU kernel
K2 to the same stream. The CPU returns immediately after the operations are
enqueued and is not blocked on the completion of the enqueued operations.
It is the GPU SEC responsibility to
\begin{tiny}\circled[text=white,fill=black,draw=black]{1}\end{tiny} launch K1
and \begin{tiny}\circled[text=white,fill=black,draw=black]{2}\end{tiny} wait for
its completion. Once K1 completes,
\begin{tiny}\circled[text=white,fill=black,draw=black]{3}\end{tiny} GPU
SEC triggers the execution of MPI operations and
\begin{tiny}\circled[text=white,fill=black,draw=black]{4}\end{tiny} waits for
these operations to finish. Next,
\begin{tiny}\circled[text=white,fill=black,draw=black]{5}\end{tiny} the GPU
SEC launches K2.

With the proposed solution, an application process running on the CPU enqueues
operations to the NIC command queue and the GPU stream, but does not get
directly involved in the control paths of MPI communication operations,
subsequent kernel launch, and tear-down operations.
The CPU does not directly wait for MPI communication operations to complete.
The GPU manages the control paths and eliminates potential synchronization
points in the application.

%% file: src/content/background.tex

\section{Triggered Operations Background}\label{sec:background}
This section describes the core components required in implementing GPU ST
communications.
Triggered operations~\cite{quadrics-tops,Beecroft2004QsNetI,tops-portal,offload-coll,tops-coll}
originally introduced in QsNet~\cite{Petrini2002TheQN}
and Portals~\cite{portals} are currently supported in several modern
network interconnects like \textbf{HPE Slingshot}~\cite{slingshot2,slingshot1,slingshot}.
This feature allows the application process from the
software layer like MPI to enqueue NIC command descriptors ahead of time in
the NIC command queue. Unlike regular NIC command descriptors, triggered
operations offer deferred execution semantics, where the execution of these
operations can be triggered at a later point in time. A modern HPC interconnect
with support for triggered operations is used to implement and explore the
proposed ST scheme. This hardware feature from the NIC is exposed to MPI through
network middleware libraries like Libfabric~\cite{libfabrics,fi_trigger}.

\subsection{Trigger Events}\label{subsec:trigger-events}
\begin{figure}[!ht]
  \includegraphics[width=\linewidth]{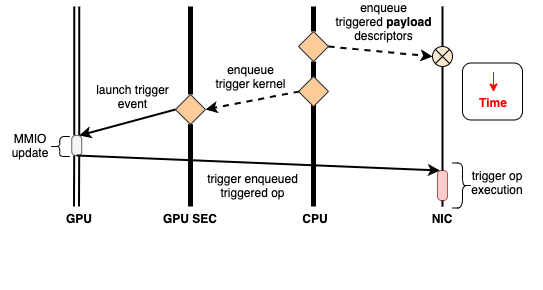}
  \caption{Trigger event interaction with GPU stream.}
  \label{fig:st-trigger}
\end{figure}

This section describes the events required to trigger execution of an enqueued
triggered operation from the GPU. The NIC command descriptor for the triggered
operations takes three parameters~\cite{fi_trigger} -
\textit{trigger counter}, \textit{completion counter}, and
\textit{trigger threshold value}.
Following are the options available to trigger the execution of an enqueued
triggered operation.
\begin{enumerate}
    \item Update the associated trigger counter to a previously defined
    threshold value; or
    \item Associate a NIC counter with a Memory Mapped I/O (MMIO)
    register. When a \textit{local store 
    operation} is performed to this
    MMIO register, it increments the register value. When the value reaches a
    previously defined threshold, it triggers the execution of the enqueued
    operation.
\end{enumerate}
For ST, a GPU triggers the execution of these NIC descriptors through the
execution of a GPU kernel. The GPU kernel does not have access to the NIC HW
counters; instead it performs a store to the MMIO register to trigger the
execution of the enqueued operation. Figure~\ref{fig:st-trigger} represents a
sequence of events involved in offloading the trigger event to the GPU SEC, and
the GPU SEC launching the GPU kernel to update the MMIO register associated
with an enqueued triggered operation.

\subsection{Completion Events}\label{subsec:completion-events}
Completions of the executed triggered operations are identified by polling the
associated completion counters from the NIC command descriptors.
Section~\ref{subsec:trigger-events} describes using a lightweight MMIO
register update to increment the trigger counter and execute the
enqueued triggered operations from the GPU. There are no such options
available for the GPU to directly poll for completions on the NIC counters.
Instead, a triggered \textit{chaining logic} is used to monitor completions.

\begin{figure}[!ht]
  \includegraphics[width=\linewidth]{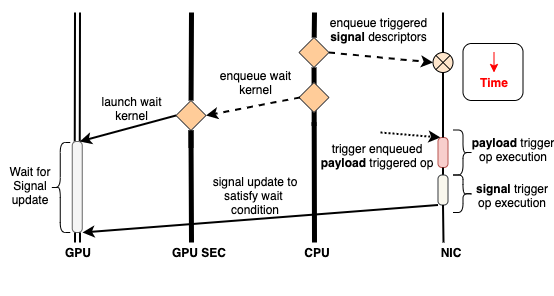}
  \caption{Completion event interaction with GPU stream.}
  \label{fig:st-complete}
\end{figure}

In the chaining approach, an enqueued triggered operation used for payload
data transfer is chained with an another triggered operation as a signaling
update. The signaling triggered operations are triggered atomic increments
updating a local GPU memory buffer. Chaining is performed by using the completion
counter specified in the payload's NIC command descriptor as a trigger counter for the
signal's NIC command descriptor. With this mapping, completion of a payload
triggered operation will automatically trigger the execution of the signal.
Figure~\ref{fig:st-complete} describes using an enqueued GPU kernel to wait for
a signal increment update on a GPU buffer from executing a signal triggered
operation chained with the payload triggered operation.

%% file: src/content/proposal.tex

\section{Stream-Triggered Communication}\label{sec:proposal}
This section introduces an MPI active-RMA-based solution to support ST
with the following semantic goals.

\subsubsection{\textbf{Support for Automatic Progress}}\label{auto-progress}
ST operations are enqueued by the CPU into the respective execution queues and
return immediately. All enqueued operations are expected to guarantee
automatic progress without blocking the CPU on the execution of the enqueued
operations.

\subsubsection{\textbf{Support for Non-blocking Communication}}\label{nbi}
All ST operations support non-blocking semantics, but these semantics are
different from traditional MPI non-blocking semantics. In the traditional
non-blocking operation, return from the operation restricts any further updates to
the memory buffers associated with the operation. Any updates to the associated
buffers can result in undefined behavior. In ST, while the CPU is restricted from
accessing the associated buffers, the GPU is free to make any updates until the
execution of the trigger event by the GPU SEC.

\subsubsection{\textbf{Support for Aggregated Communication}}\label{aggregate}
Hardware resources used to offload ST operations (like the NIC counters) 
are limited. While different implementation design options are used to
efficiently manage these resources, it is critical for the proposed
solution to effectively support them. Aggregation allows multiple ST
operations to be batched and use fewer resources for offloading
communications.

The proposed MPI active-RMA-based solution provides support for all the above
detailed expected semantics. Description of the proposal is provided
below.

\subsection{MPI RMA Overview}\label{subsec:rma-overview}
\begin{figure}[!ht]
  \includegraphics[width=0.6\linewidth]{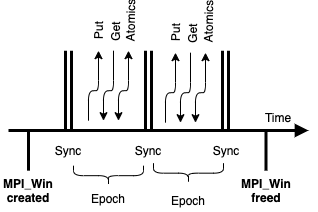}
  \caption{Typical MPI window timeline.}
  \label{fig:rma-overview}
\end{figure}

The MPI RMA communication model supports using (1) window creation for exposing
remote accessible memory locations, (2) one-sided data movement operations, and
(3) synchronization operations. Figure~\ref{fig:rma-overview} represents the
timeline of window creation, calls to RMA data movement, and synchronization
operations.

\begin{enumerate}
    \item Creation of \emph{MPI\_Win} establishes a memory window.
    \item Synchronization is required to ensure memory consistency and coherence,
    as all RMA data movement operations offer non-blocking semantics.
    \item RMA epochs are defined as the execution span within synchronization
    operations. The specific calls used for synchronization depend on whether
    the active or passive target synchronization model is being used. MPI RMA
    data movement operations are performed within an epoch.
    \item Calls to \emph{MPI\_Put}, \emph{MPI\_Get}, and \emph{MPI\_Accumulate}
    RMA data movement operations must be contained within an epoch.
    Multiple data transfers can occur within the same epoch,
    amortizing the performance cost of synchronization operations.
\end{enumerate}

\subsection{MPI Active RMA Overview}\label{subsec:arma-overview}
In the active target synchronization paradigm, both the origin and target
processes are actively involved in the synchronization operations. An RMA
synchronization operation at the target process starts the \textit{exposure
epoch} that allows the origin processes to start accessing the memory window. A
synchronization operation at the origin process starts the \textit{access epoch}
that allows it to issue data movement operations on the remotely accessible
memory window.

\begin{figure}[!ht]
  \includegraphics[width=0.85\linewidth]{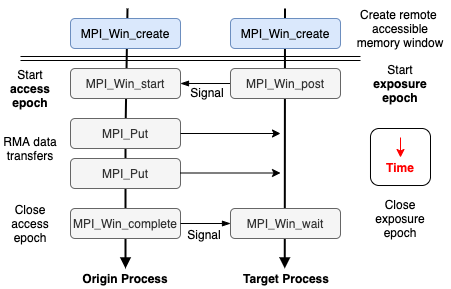}
  \caption{MPI active RMA using post-wait-start-complete.}
  \label{fig:arma-overview}
\end{figure}

Figure~\ref{fig:arma-overview} demonstrates MPI active RMA by one origin and
one target process using \emph{MPI\_Win\_post}, \emph{MPI\_Win\_start},
\emph{MPI\_Win\_complete}, and \emph{MPI\_Win\_wait} operations. Initially a
remotely accessible memory window on all participating processes in the MPI
communicator is created using \emph{MPI\_Win\_create} operation. With the
\emph{MPI\_Win} object created, Figure~\ref{fig:arma-overview} represents the
the exposure and access epochs, and the multiple one-sided data
transfer operations performed within them.

Availability of the exposure epoch on the target process, and the
intent to close the access epoch on the origin process are performed by sending
signals internally in the MPI runtime. \emph{MPI\_Win\_post} generates signals
on opening the exposure epoch, and the \emph{MPI\_Win\_start} waits for these
signals to arrive. Similarly, \emph{MPI\_Win\_complete} initiates signals
after completing all data transfer operations, and \emph{MPI\_Win\_wait} waits
for the arrival of all these completion signals.

\subsection{Motivation for Active RMA Usage}\label{subsec:arma-motivation}
Multiple API options were considered for introducing GPU stream-awareness in
MPI. Apart from providing support for the expected semantics from
Sections~\ref{auto-progress}-~\ref{aggregate}, MPI active RMA operations
support the following semantics that motivates using the active RMA
communication model for introducing GPU Stream-awareness.

RMA data movement operations in general \emph{do not} use message matching
~\cite{hw_mm,bmm,keith-message-matching,Flajslik2016MitigatingMM,Bayatpour2016AdaptiveAD,Schonbein2018MeasuringMM,osti_hw_mm}
semantics for message ordering as supported by traditional MPI point-to-point
(P2P) communication operations. This is critical for offloading the
intra-node ST operations into the GPU kernel. Implementing message matching
semantics for intra-node operations requires using a progress thread to emulate
the required deferred execution ST semantics. Using an RMA based communication
model eliminates this implementation difficulty.

\begin{figure}[!ht]
  \includegraphics[width=0.85\linewidth]{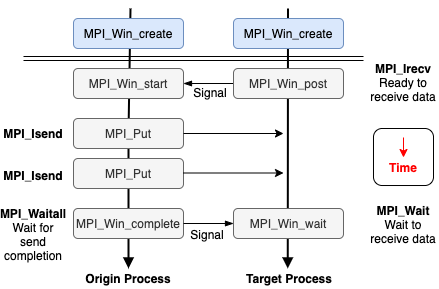}
  \caption{Comparing MPI P2P and MPI active RMA.}
  \label{fig:arma-compare}
\end{figure}

While both active and passive RMA models support the necessary semantics, the
ease of application conversion using the existing MPI P2P communication is an
another major reason for active RMA selection. Figure~\ref{fig:arma-compare}
shows the similarities in the semantics for MPI active RMA with the traditional
MPI P2P operations. As shown in Figure~\ref{fig:arma-compare}, most MPI P2P
operations have an equivalent matching operation in active RMA. With many GPU
aware applications using MPI P2P operations, providing support for GPU
stream-awareness through MPI active RMA allows the application programmers to
port to the new proposed APIs.

\subsection{GPU Stream-Triggered Active RMA}\label{subsec:st-arma}
GPU stream-awareness in active RMA is proposed by introducing new MPI operations.
Summary of the new proposed operations is provided below.
\begin{enumerate}
    \item \emph{MPIX\_Win\_post\_stream}, \emph{MPIX\_Win\_complete\_stream},
    and \emph{MPIX\_Win\_wait\_stream} take a GPU stream argument as input and
    replace existing \emph{MPI\_Win\_post}, \emph{MPI\_Win\_complete}, and
    \emph{MPI\_Win\_wait} active RMA operations.
    \item Proposed GPU stream aware operations are used to enqueue (a) the
    signals internally used by the runtime to create and manage the epochs, and
    (b) the data transfer operations into the GPU and NIC execution queue. An
    application process returns immediately after enqueuing operations.
    \item Proposed operations allow offloading both signals and data transfers
    from the application host process to the GPU. These
    operations are non-blocking with respect to the application host process.
\end{enumerate}
Along with managing compute kernels defined by the application, the proposed ST
scheme enables the GPU to coordinate with the NIC in managing
the control paths of MPI operations.

\subsection{Stream-Triggered Active RMA Usage Model}\label{subsec:st-arma-usage}
\begin{figure}[!ht]
  \includegraphics[width=\linewidth]{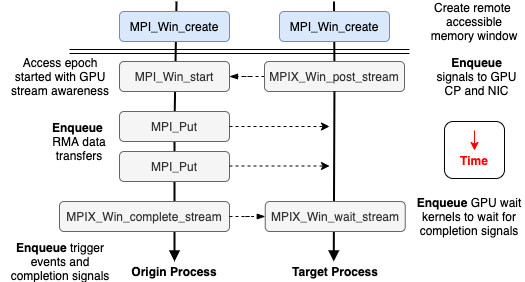}
  \caption{MPI ST active RMA usage model.}
  \label{fig:st-arma-usage}
\end{figure}

Figure~\ref{fig:st-arma-usage} represents the sequence of operations supported
by the ST active RMA. From the application process perspective, all
synchronization (post-start-complete-wait) and data transfer operations are
non-blocking.

\begin{enumerate}
    \item \emph{MPIX\_Win\_post\_stream} enqueues signals and returns
    immediately. Once returned from the operation, the application process is
    free to execute other operations. All enqueued signals are executed in-order
    by the GPU stream.
    \item \emph{MPIX\_MODE\_STREAM} is a new RMA execution mode introduced as
    part of the proposed change. Using this mode with \emph{MPI\_Win\_start}
    notifies the MPI runtime that subsequent data movement operations executed,
    like the \emph{MPI\_Put} as shown in Figure~\ref{fig:st-arma-usage},
    at the origin process are GPU stream-aware. These data movement operations
    are just enqueued and not executed immediately by the host process.
    \item \emph{MPIX\_Win\_complete\_stream} enqueues both trigger events and
    completion signals for previously enqueued payload data transfer operations.
    \item \emph{MPIX\_Win\_wait\_stream} enqueues GPU wait kernels.
    The GPU wait kernels poll for the completion signal
    updates from the origin process. The host process enqueues the GPU wait
    kernels and returns immediately.
\end{enumerate}
Overall from the application host process perspective, the proposed ST active
RMA solution enqueues signals and data transfer events to the GPU and the NIC
execution stream, and returns immediately. And the control path of the operations
is offloaded into the GPU. This property satisfies the major design goal for
minimizing the need for the CPU cycles to drive the control paths of the MPI
communication operations.

\subsection{Stream-Triggered Active RMA Example}\label{subsec:st-arma-example}
Figure~\ref{fig:example} provides a simple example to describe the functionality
of the proposed ST active RMA operations and compares them to regular active RMA
operations. In this example, the window and group creation operations are not
shown. Instead, only the core communication and synchronization operations are
discussed.

Every process in the example acts both as an origin and as a target. Each
process writes into the remote accessible target memory location on its left and
right ranks, and expects updates from its left and right ranks. As shown in
Figure~\ref{fig:baseline-example}, each process opens its exposure epoch
using \emph{MPI\_Win\_post}. This is followed by a kernel launch of \emph{increment}
to update the memory location that the process is preparing to transfer to its
left and right neighbors. Before using the \emph{src} buffer for any further
RMA operation (like \emph{MPI\_Put}) by opening an access epoch using
\emph{MPI\_Win\_start}, the application process is required to synchronize with
the GPU stream to guarantee the availability of the \emph{src} buffer.
\emph{hipStreamSynchronize} is used for synchronization and has non-negligible
performance cost.

\emph{MPI\_Win\_complete} ensures the completion of all posted RMA
operations in the access epoch. \emph{MPI\_Win\_wait} is used to wait for the
closure signal from the access epoch to close its exposure epoch. With
the exposure epoch closed, all received data from its left and right neighbors
are accessed using \emph{compare} kernels.

The example in Figure~\ref{fig:baseline-example} shows multiple iterations. A
call to synchronize with the GPU stream using \emph{hipStreamSynchronize} is
required to ensure the completion of all operations before proceeding to the
next iteration. Synchronizing the application process with the GPU kernel has
non-negligible performance costs. Two synchronization calls are required:
\begin{enumerate}
    \item To validate the availability of the \emph{src} buffers updated by the
    \emph{increment} GPU kernel for data transfer operations in the access
    epoch, and
    \item To validate the completion of the verification GPU kernel
    \emph{compare} and reuse the received buffer for next iteration.
\end{enumerate}
Figure~\ref{fig:st-example} shows the ST variant of the example shown
in Figure~\ref{fig:baseline-example}. Since calls to
\emph{MPIX\_Win\_post\_stream}, \emph{MPIX\_Win\_complete\_stream}, and
\emph{MPIX\_Win\_wait\_stream} are non-blocking with respect to the application
process, all operations are enqueued to the required execution queues, and the
application host process returns without being blocked on the completion of
execution of the enqueued operations. Implicit in-order execution
guarantees by the GPU stream allows enqueuing multiple iterations of the
operations without synchronizing the application process with the GPU stream.
Usage of a single \emph{hipStreamSynchronize} call at the end of all the
iterations is shown in Figure~\ref{fig:st-example}.

\begin{figure}[!ht]
\centering
\begin{subfigure}{0.48\textwidth}
\lstset{
    language=C,
    basicstyle=\tt\small,
    frame = single,
    morekeywords={MPI_Win_start, MPI_Put, MPI_Win_complete, MPI_Win_wait,
    hipStreamSynchronize, MPI_Win_post},
    moredelim=**[is][\color{red}]{@}{@}
}
\begin{lstlisting}
for (n = 0 -> niter) {
    MPI_Win_post(group, 0, win);

    increment<<<nb,nt,0,stream>>>(n, src);
    @hipStreamSynchronize(stream);@
    MPI_Win_start(group, 0, win);
    MPI_Put(src,...left_rank,..., win);
    MPI_Put(src,...right_rank,..., win);
    MPI_Win_complete(win);

    MPI_Win_wait(win);
    compare<<<nb,nt,0,stream>>>(n,from_left_rank);
    compare<<<nb,nt,0,stream>>>(n,from_right_rank);
    @hipStreamSynchronize(stream);@
}
\end{lstlisting}
\caption{Baseline MPI program with active RMA communication.}
\label{fig:baseline-example}
\end{subfigure}
\begin{subfigure}{0.48\textwidth}
\lstset{
    language=C,
    basicstyle=\tt\small,
    frame = single,
    moredelim=**[is][\color{red}]{@}{@},
    morekeywords={MPI_Win_start, MPI_Put, MPIX_Win_post_stream,
    MPIX_Win_complete_stream, MPIX_Win_wait_stream,
    hipStreamSynchronize},
}
\begin{lstlisting}
for (n = 0 -> niter) {
    MPIX_Win_post_stream(group, 0, win);

    increment<<<nb,nt,0,stream>>>(n, src);
    MPI_Win_start(group, 0, win);
    MPI_Put(src,...left_rank,..., win);
    MPI_Put(src,...right_rank,..., win);
    MPIX_Win_complete_stream(win);

    MPIX_Win_wait_stream(win);
    compare<<<nb,nt,0,stream>>>(n,from_left_rank);
    compare<<<nb,nt,0,stream>>>(n,from_right_rank);
}
@hipStreamSynchronize(stream);@
\end{lstlisting}
\caption{MPI program with GPU ST active RMA communication.}
\label{fig:st-example}
\end{subfigure}
\caption{Compare ST and baseline MPI programs.}
\label{fig:example}
\end{figure}

The example shown in Figure~\ref{fig:st-example} shows the application
process fully offloading the MPI control path to the GPU. Enqueuing multiple
iterations of operations into the GPU and synchronizing with
the GPU stream at the end of all iterations shows the offloading semantics of
the ST active RMA.

%% file: src/content/implementation.tex

\section{ST Active RMA Implementation}\label{sec:st-arma-implementation}
This section provides the ST implementation details.

\subsection{Inter-node Communication}\label{subsec:st-arma-inter-node}
Inter-node ST active RMA uses the NIC-based triggered operations for
implementing the required deferred execution semantics. Triggered operations
and associated trigger and completion events were discussed in
Section~\ref{subsec:trigger-events}.

\subsubsection{\textbf{Triggered Operations Usage from Origin Process}}
\label{subsubsec:inter-origin}
\begin{figure}[!ht]
  \includegraphics[width=0.75\linewidth]{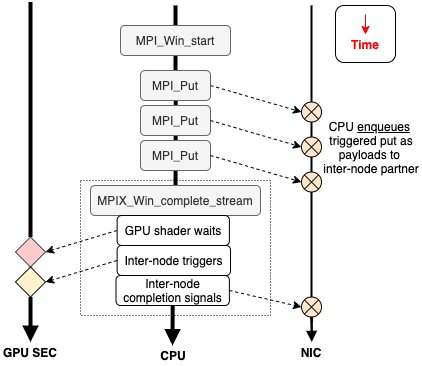}
  \caption{Triggered operations usage from origin process.}
  \label{fig:tops-usage-origin}
\end{figure}

Figure~\ref{fig:tops-usage-origin} represents the triggered
operations usage from the origin process perspective.
\begin{enumerate}
    \item When \emph{MPI\_Win\_start} is used with \emph{MPIX\_MODE\_STREAM},
    the MPI runtime updates the internal metadata for the MPI window used
    in the operation to be GPU stream aware. This update gets released only by
    closing the access epoch using \emph{MPIX\_Win\_complete\_stream}.
    \item In the example shown in Figure~\ref{fig:tops-usage-origin}, within
    the access epoch there are three inter-node data transfers using
    \emph{MPI\_Put}. Separate triggered operation descriptors are enqueued per
    \emph{MPI\_Put} to the NIC command queue. The application process returns
    immediately after the enqueue operations. All the enqueued triggered
    operations within an access epoch use the same threshold value, same
    triggering, and same completion counter.
    \item \emph{MPIX\_Win\_complete\_stream} handles opening the access epoch on
    the GPU execution stream, enqueues a trigger event for all the previously
    enqueued triggered data transfers, and signals the target process on the
    completion of the access epoch.
    \begin{itemize}
        \item Initially a GPU kernel is enqueued into the GPU execution stream
        and the CPU returns immediately. The GPU kernel polls for the signal
        update from the target process on the availability of the exposure
        epoch. All further operations on the GPU stream are blocked, and the only
        option to successfully exit the kernel execution is the arrival of the
        signaling updates from the target process.
        \item A GPU kernel to update the NIC memory mapped I/O register
        associated with the triggering counter used in all the previously
        enqueued triggered operations is enqueued.
        \item New triggered operations are enqueued into the NIC command queue
        as completion signals to all target processes in the window. The
        completion counter used in all the previously enqueued triggered
        operations is now used as a trigger counter for all the signaling
        triggered operations.
    \end{itemize}
\end{enumerate}

\subsubsection{\textbf{Triggered Operations Usage from Target Process}}
\label{subsubsec:inter-target}
\begin{figure}[!ht]
  \includegraphics[width=\linewidth]{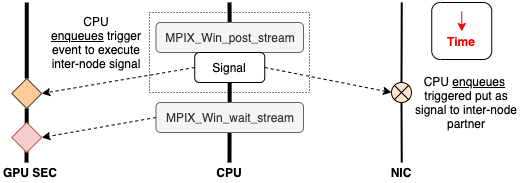}
  \caption{Triggered operations usage from target process.}
  \label{fig:tops-usage-target}
\end{figure}

Figure~\ref{fig:tops-usage-target} represents the triggered
operations usage from the target process perspective.
\begin{enumerate}
    \item When the application process calls \emph{MPIX\_Win\_post\_stream} to
    initiate the start of the exposure epoch, signals are sent to all origin
    processes in the window. Signals enqueued are triggered inter-node Put
    operations. Corresponding trigger events used to trigger the enqueued
    signals are also immediately enqueued into the GPU SEC execution queue.
    \item The \emph{MPIX\_Win\_wait\_stream} operation enqueues a GPU kernel in
    the GPU SEC execution queue to wait on the arrival of the completion signals
    from the origin process. Wait operations are GPU kernels polling on the
    memory locations expecting a signal update.
\end{enumerate}

Overall, the triggered operations used on both the origin and the target
process are stateless. Triggered operations usage shows the option to fully
offload the control path of the MPI RMA operations into the GPU. The
application process is free to return from all the operations after enqueuing
the required triggered operations, and the trigger and wait GPU kernels.
As described in Section~\ref{subsec:streams}, in-order execution of all enqueued
operations in the GPU stream is guaranteed by GPU SEC.

\subsection{Triggered Operations Throttling}\label{subsec:throttling}
As described in Section~\ref{subsec:st-arma-inter-node}, all inter-node ST
operations are fully offloaded as NIC triggered operations. But
NIC triggered operation resources are limited. It is not possible to
indefinitely enqueue triggered operations into the NIC. The CPU will eventually be
blocked and needs to synchronize with the NIC to recapture the triggered
operation resources for re-use. This section describes the different
throttling algorithms designed to manage the resources.

\subsubsection{\textbf{Application-Level Throttling}} Application-level
throttling introduces synchronization points in the application to avoid
over-subscription. This is similar to the standard active RMA model without
GPU stream-awareness. The application process is synchronized with the GPU
stream (using operations like \emph{hipStreamSynchronize}) for every few
iterations and waits for all the previous iterations to complete before
proceeding further.

Functionally this scheme moves the throttling requirement from the MPI runtime to the
application programmer. Though implementable, the practical usability of this
approach is limited. It is hard for applications to determine either the synchronization
points or the number of triggered operations internally generated by the ST
MPI runtime per iteration. The limitations of this approach motivate the need
for other throttling algorithms, and this approach is mostly used to introduce
the need for throttling in the MPI ST runtime.

\subsubsection{\textbf{Static Throttling}} Static throttling converts the
sync point described in the application-level throttling into weak
sync points within the ST MPI implementation. Instead of using heavy
options like \emph{hipStreamSynchronize}, in this scheme the ST implementation
uses the CPU to poll for completion of all previously posted triggered
operations and determine the availability of triggered operation resources
before attempting to enqueue any new operations.

The core semantic of this scheme is to wait for completions of all previously
posted operations before enqueuing any new operations. This completion counter
evaluation is performed on an epoch basis, instead of every triggered operation.
An epoch may consists of multiple triggered operations.

\subsubsection{\textbf{Adaptive Throttling}} Adaptive throttling is an
incremental update from the static throttling. Similar to the static throttling,
this scheme is implemented in the ST implementation and not in the application
directly.
The main difference is that in adaptive throttling, instead of waiting for the
completions of all previously posted operations, triggered resources are
recaptured and immediately used as soon as the previous operations are
completed. As the name suggests, the adaptive throttling algorithm adapts to the
available number of triggered resources in the NIC and recaptures them for
reuse as soon as they are available for ST active RMA usage.

\subsection{Intra-node Communication}\label{subsec:st-arma-intra-node}
This section briefly describes the intra-node ST implementation. Three different
types of GPU kernels are used to implement the intra-node sequence of events
required to implement ST active RMA:
\begin{enumerate}
    \item GPU kernels performing signal updates,
    \item GPU kernels performing payload data transfers, and
    \item GPU kernels performing wait operations on the incoming signals.
\end{enumerate}
The three different GPU kernels used are accessing MPI RMA window memory exposed
by other MPI ranks using GPU IPC-based~\cite{CUDA-IPC} transfers. The GPU signaling
kernels perform a simple store operation onto the signal buffers on the other
MPI ranks' part of the MPI window. Payload transfers perform memory copy
operations on the target process memory, which is part of a separate process
address space, using IPC communications. The wait kernels are polling on a local
memory location expecting signal updates from all ranks.

In brief, the sequence of events for intra-node operations is similar to the
inter-node logic described in Section~\ref{subsubsec:inter-origin} and
Section~\ref{subsubsec:inter-target}. The major difference is that all
triggered operations from the inter-node design are replaced with the above
described GPU kernels, and the kernels are enqueued into the GPU execution
stream. All triggered operations for signaling and payload transfers from the
inter-node design are replaced with GPU kernels performing signal updates and
payload memory copy operations using GPU IPC communication.

Similar to the inter-node design, intra-node ST transfers are also fully
offloaded into the GPU, and the implementation is stateless. This intra-node
implementation is feasible, as the MPI RMA semantics do not require any
message matching~\cite{comb} based ordering semantics.

\subsection{Merged GPU Kernels}\label{subsec:merged-gpu}
If the MPI group used in the synchronization operations consists of multiple MPI
ranks, signals must be generated for all the participating processes. Launching
a GPU kernel per signaling operation is expensive. As all the GPU kernels are
generated either from the operation \emph{MPIX\_Win\_complete\_stream} or
\emph{MPIX\_Win\_post\_stream}, it is possible for the MPI runtime to aggregate
the launch of all operations as a single merged kernel updating multiple
different locations. Similar to using a merged signaling kernel, it is possible
to independently merge the wait and memory copy kernels used for intra-node
ST operations.
Optimized implementations of the above described merged kernels are designed
specifically for the ST implementation, and the performance benefits of these
kernels are evaluated in Section~\ref{subsec:merged-gpus}.

%% file: src/content/performance.tex

\section{Performance Analysis}\label{sec:performance}
In this section, the Faces microbenchmark is used to evaluate the
performance of the prototype ST MPI communication operations.

\subsection{Test System}\label{subsec:test-system}
All tests were performed on a system based on a node architecture consisting of
AMD CPUs and GPUs. Each node offers a single AMD Milan CPU socket, four MI200
GPU modules, and four HPE Slingshot NICs~\cite{slingshot2,slingshot1,slingshot}.
Each GPU module consists of two GPU devices (GCDs).
Each NIC is attached to a single MI200 module via PCIe Gen4. The MI200 modules
are interconnected with 1x or 2x high speed xGMI3 links. A prototype
implementation of ST active RMA was used for all experiments, along with
ROCm~\cite{rocr}, and Libfabric~\cite{libfabrics} for the GPU and NIC interface.

\subsection{Test Case Overview}\label{subsec:test-case}
The Faces microbenchmark is inspired by the nearest-neighbor communication pattern
from the CORAL-2~\cite{coral2-bm} Nekbone~\cite{nekbone-bm} benchmark. Faces
represents a common communication pattern from multiple user applications. Faces
communicates with at most 26 neighbors and transfers the faces, edges,
and corners of spectral elements on the surface of the local block. Weak scaling
Faces has three nested loops that performs the following: (1) \textbf{outer
loop} allocates MPI windows, runs loops, and deallocates windows, (2)
\textbf{middle loop} initializes the values of the spectral elements and runs the
inner loop; and (3) \textbf{inner loop} runs the communication steps and
accumulates the wall-clock runtime.
Three different versions of the Faces benchmark are used for this analysis:
\begin{itemize}
    \item Traditional P2P Faces using MPI P2P communication;
    \item Active RMA Faces using standard active RMA;
    \item And, ST Active RMA Faces using the proposed ST features from this work.
\end{itemize}

\subsection{Overall Faces Performance}\label{subsec:over-perf}
Results of the single and multi-node runs using the active RMA and ST
active RMA variants of the Faces benchmark are reported in
Figure~\ref{fig:overall}. The multi-node test uses 64
application processes (or MPI ranks) distributed across eight nodes, with eight
ranks per node. On each node, a one-to-one mapping between MPI ranks and GCDs is
enforced.

On average the ST active RMA version of the Faces microbenchmark shows
\textbf{23\% better performance} than the active RMA version of the Faces
microbenchmark. Figure~\ref{fig:overall} shows the execution time involved in
running the Faces benchmark. The ST active RMA implementation in this test makes
use of the optimized merged GPU kernels and optimized adaptive throttling
algorithm to fully offload the communication control path into the GPU for
all inter-node and intra-node communication operations. This performance
improvement validates the expected performance improvements involved with the
original design goal in minimizing the need for the application process running
on the CPU to drive the control paths of MPI communication operations.

\subsection{Impact of Intra-node Operations}\label{subsec:intra-node}
\begin{figure}[!ht]
  \includegraphics[width=\linewidth]{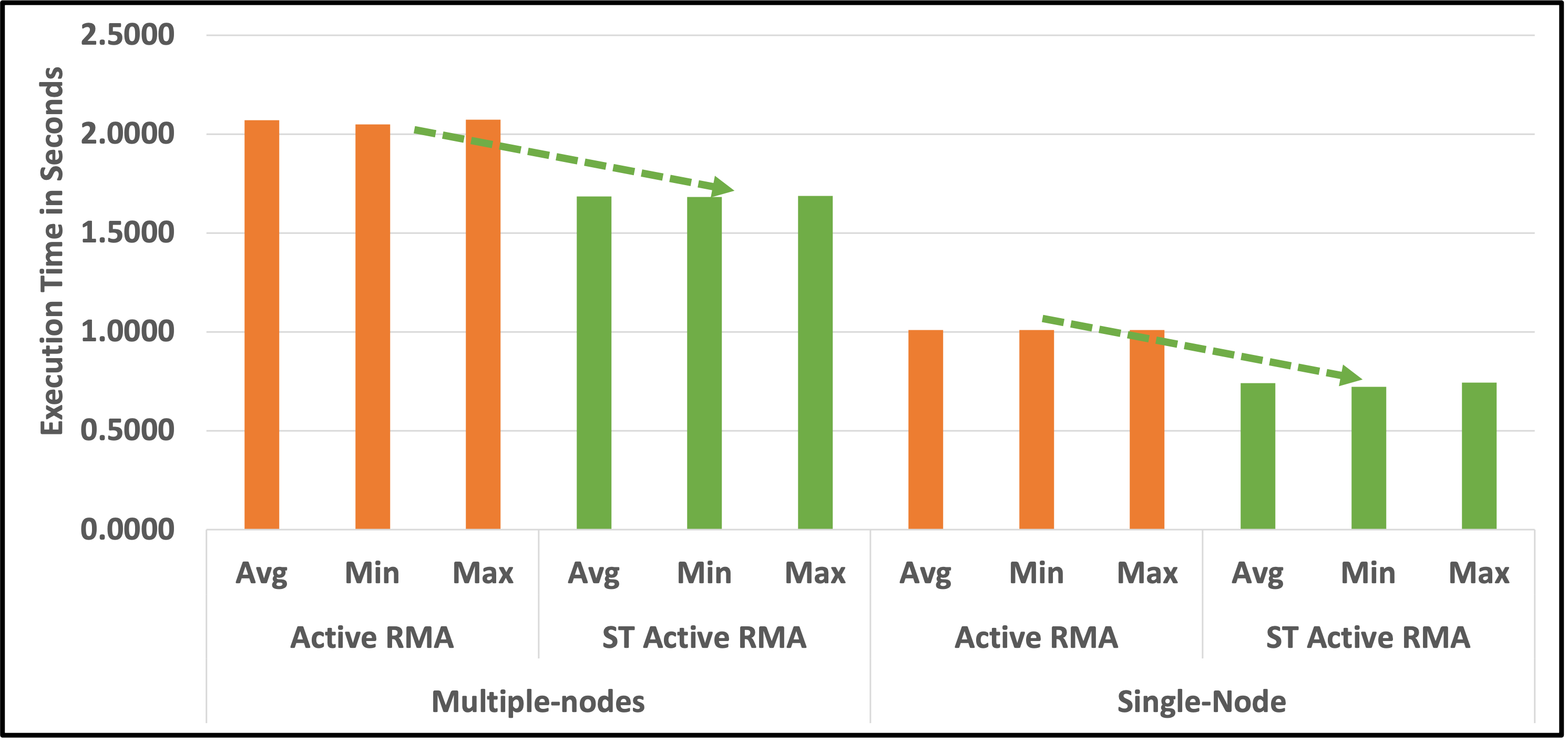}
  \caption{Faces multiple and single-node performance analysis.}
  \label{fig:overall}
\end{figure}

Results of Faces using the active RMA and ST active RMA variants within a
single node are reported in Figure~\ref{fig:overall}. The reported single-node
test uses 8 MPI ranks within a single node.
Performance of ST active RMA shows \textbf{36\% improvement} over the active RMA
variant. The ST active RMA implementation in this test makes use of the
optimized merged GPU kernels discussed in Section~\ref{subsec:merged-gpu}. No
throttling algorithm is involved, as all communication operations are within a
single node. This performance improvement validates the expected performance
improvements involved in fully offloading the intra-node communication
operations from the application process running on the CPU to the GPU.

\subsection{Impact of Throttling Triggered Operations}\label{subsec:tops}
\begin{figure}[!ht]
  \includegraphics[width=\linewidth]{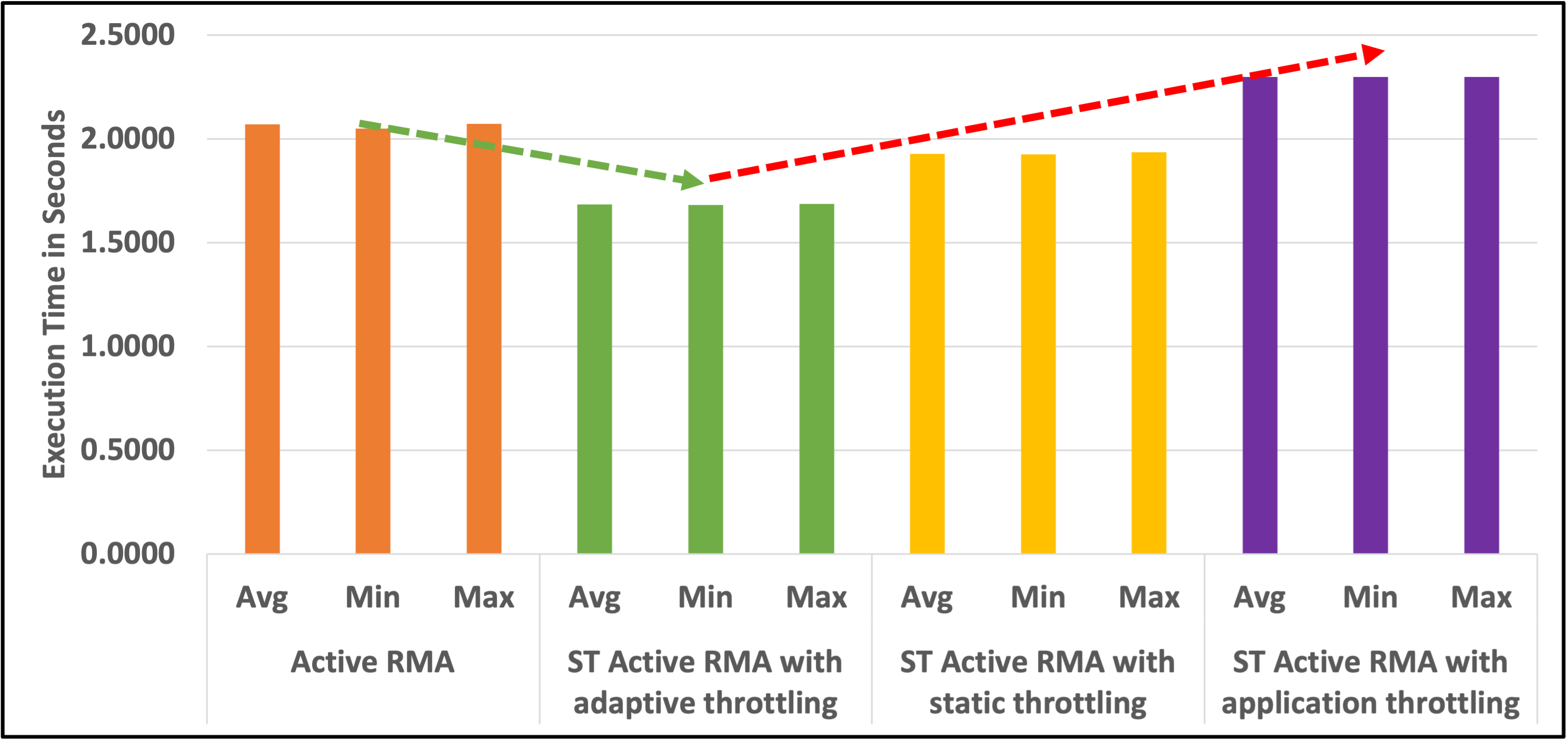}
  \caption{Impact of different throttling algorithms.}
  \label{fig:throttle}
\end{figure}

Section~\ref{subsec:throttling} describes the different algorithms used to
effectively manage the triggered operation communication resources.
Figure~\ref{fig:throttle} shows the impact of using different throttling
algorithms. In brief, for an 8-node test with 8 MPI
processes per node (64 processes total), ST active RMA using adaptive throttling shows the best
performance improvement over other throttling options. Using adaptive throttling
shows 23\% improvement over the active RMA version without GPU stream awareness.
While the performance difference between using adaptive and static throttling is
not huge, there is a 10\% improvement in using adaptive throttling. Adaptive
throttling shows 21\% improvement over application-level throttling.

\subsection{Impact of Merged GPU Kernels}\label{subsec:merged-gpus}
The ST active RMA implementation used for reporting the performance in
Figure~\ref{fig:overall} employs merged GPU kernels, as
discussed in Section~\ref{subsec:merged-gpu}. This section describes the impact
of using merged GPU kernels in the implementation. 8 nodes with 8 MPI processes
per node, and 1 node with 8 MPI processes per node are both used for this
analysis.

\begin{figure}[!ht]
  \includegraphics[width=\linewidth]{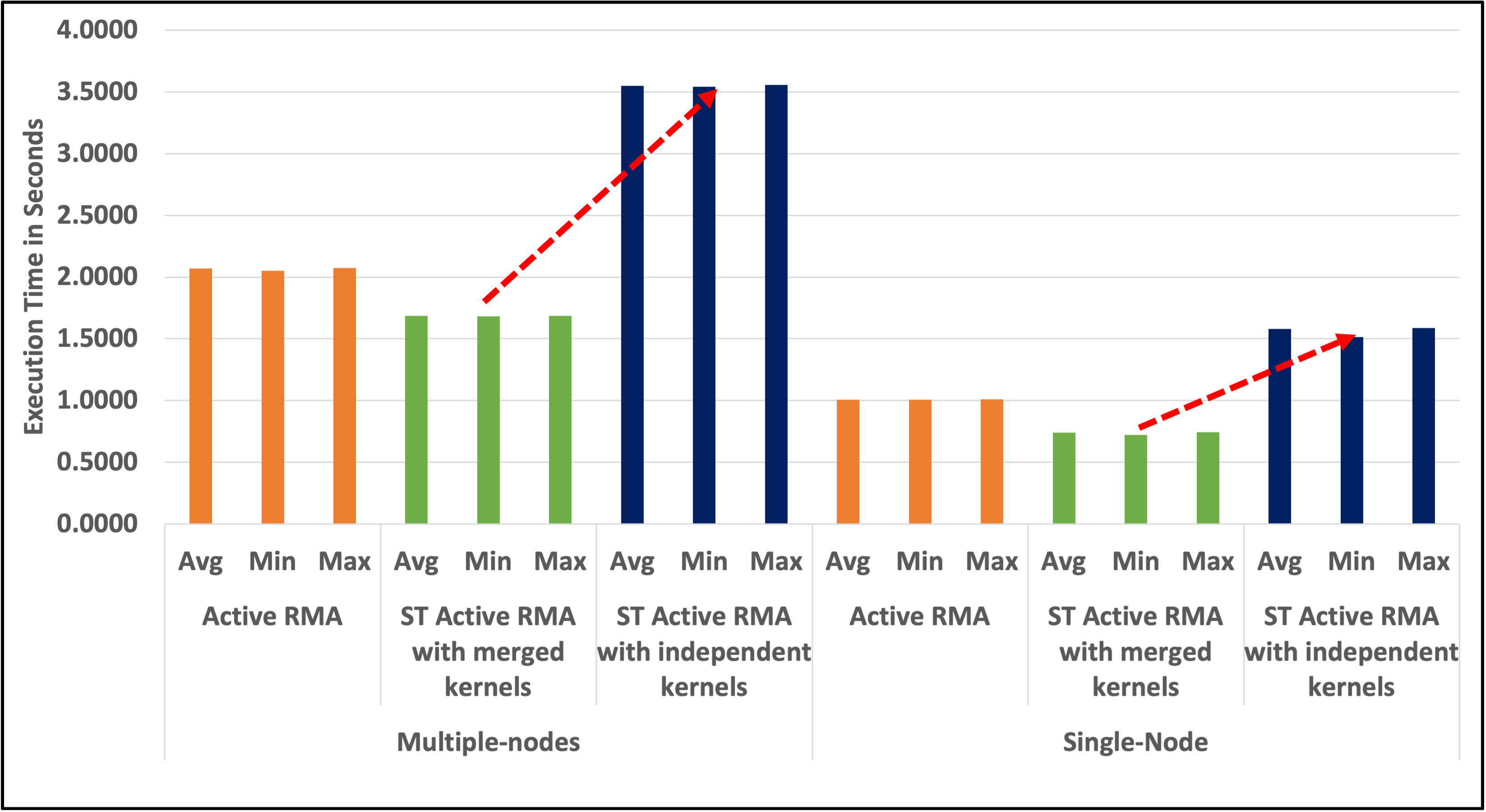}
  \caption{Impact of using merged GPU kernels.}
  \label{fig:merged}
\end{figure}

Using independent GPU kernels introduces expensive launch and cleanup
cost. Using merged GPU kernels
minimizes the number of launched GPU kernels and with that the performance cost
of the ST active RMA operations. 
Performance results from Figure~\ref{fig:merged} show the need for using merged
GPU kernels in the ST implementation.
Using merged kernel shows around 90\% and
2\emph{X} improvement over launching independent kernels across multiple nodes
and within a single node, respectively.

\subsection{Impact of Overlapping Compute Kernels}\label{subsec:overlap}
Faces can test the overlap of MPI communication with independent local
computation, in the form of a separate compute kernel launched on an independent
GPU stream. This section details the impact of adding a compute kernel in the
Faces microbenchmark. A configuration with 8 nodes and 8 MPI processes per node is
used for this test.
\begin{figure}[!ht]
  \includegraphics[width=\linewidth]{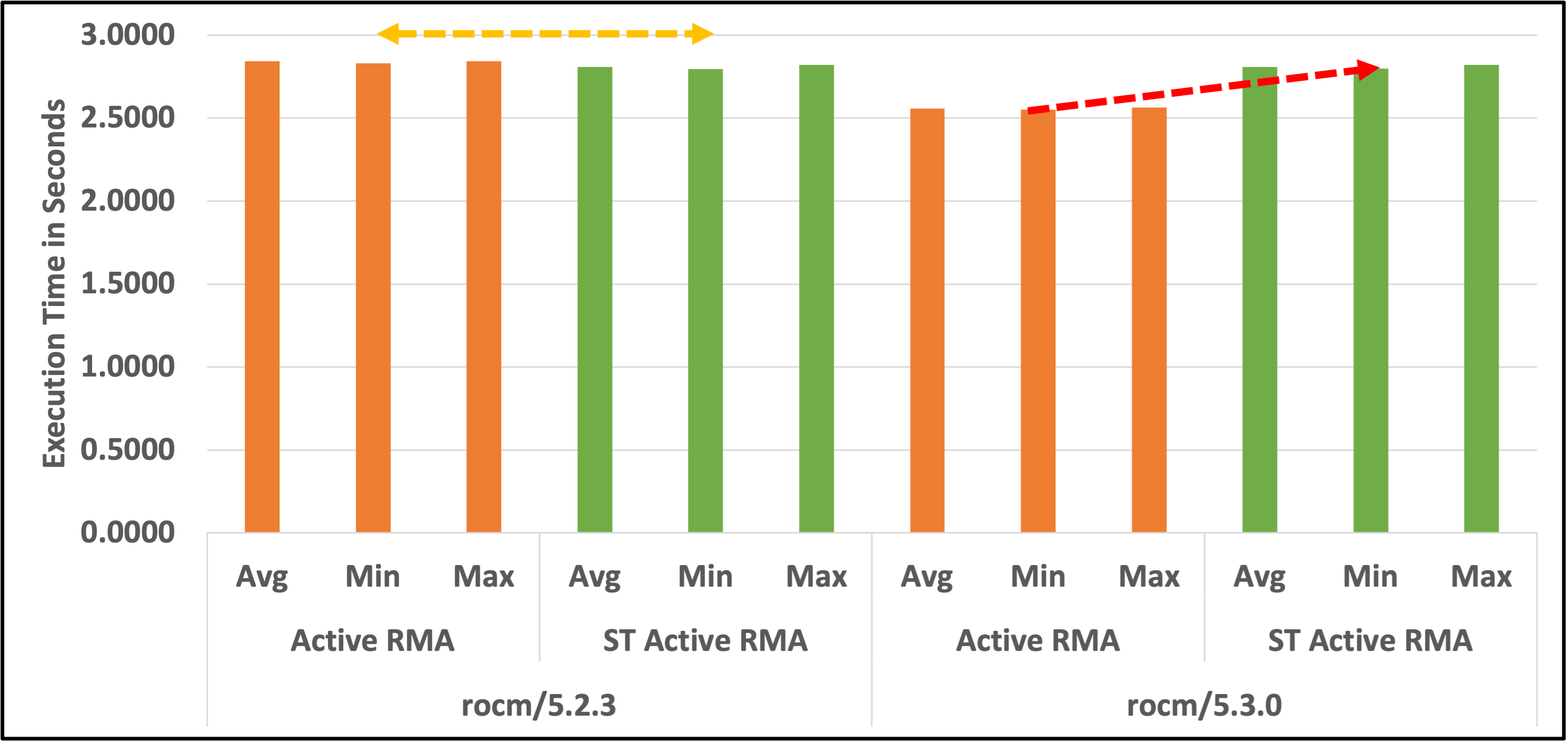}
  \caption{Impact on using overlapping compute kernels.}
  \label{fig:overlap}
\end{figure}
From Figure~\ref{fig:overlap}, the ST active RMA implementation with an
overlapping compute kernel shows at most only a 3\% performance improvement over
the active RMA variant of the benchmark without GPU stream awareness. With
changing the version of the ROCm library used, the performance is hit by 8\%.
This is significantly different from the performance behavior with just the
communication operations as discussed in Section~\ref{subsec:over-perf}.

This significant change in performance and the variations of the performance
behavior with changing the ROCm library version is not fully studied. 

\subsection{Relative Comparison with Traditional P2P}\label{subsec:p2p-perf}
\begin{figure}[!ht]
  \includegraphics[width=0.80\linewidth]{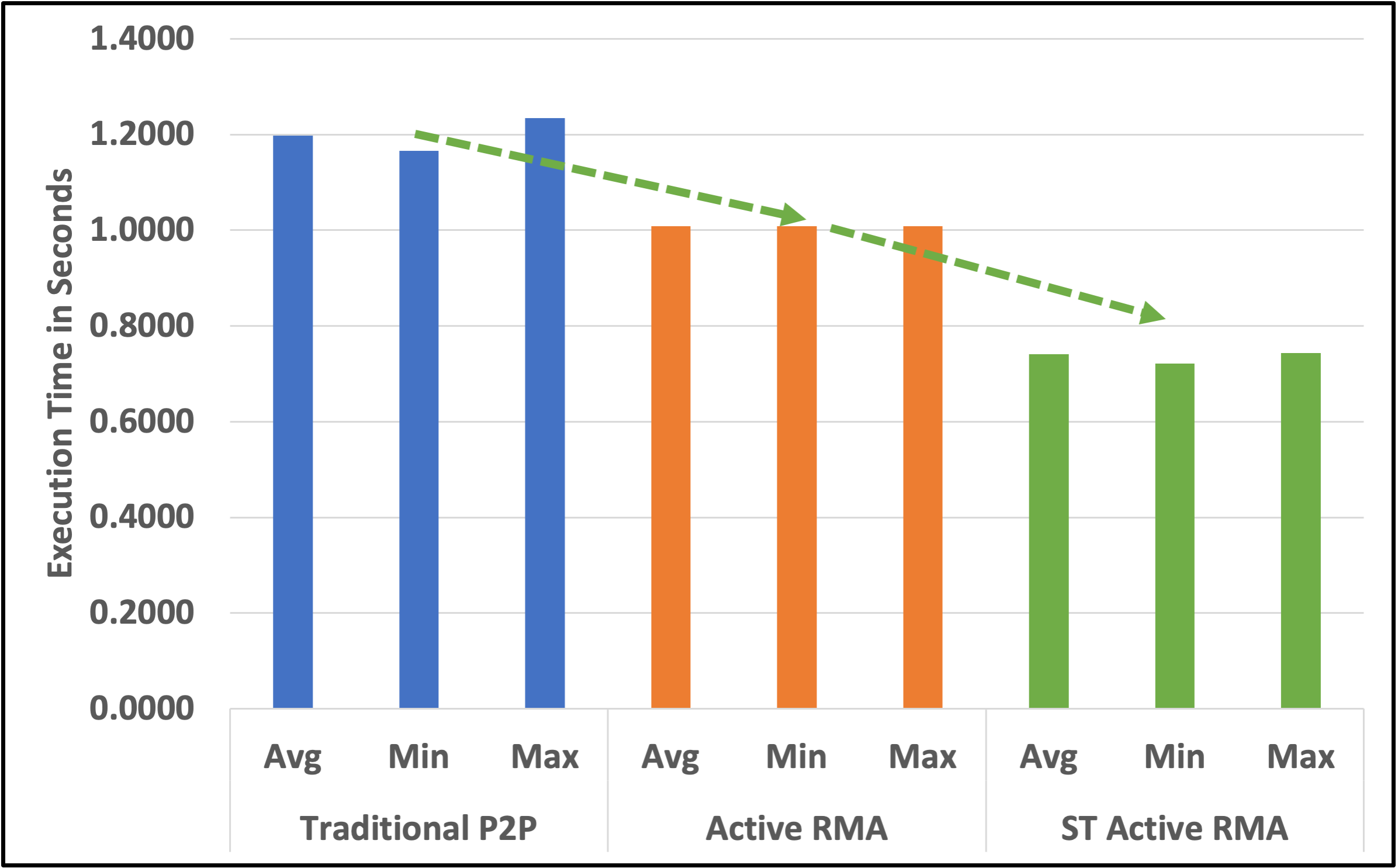}
  \caption{Intra-node comparison with traditional P2P.}
  \label{fig:intra-p2p}
\end{figure}

In this section, performance data from the traditional P2P variant of the Faces
microbenchmark is compared against the active RMA and ST variants. 
Figure~\ref{fig:intra-p2p} shows performance results for runs within a single node with 8 MPI
ranks per node. Performance of active RMA is 19\% better than the traditional
P2P implementation.
On top of the performance improvement in using the active RMA, introducing GPU
stream-awareness using the ST active RMA operations improves the performance by
36\%. 
ST thus shows 61\% improvement over the
traditional P2P variant of the Faces benchmark.

\begin{figure}[!ht]
  \includegraphics[width=0.80\linewidth]{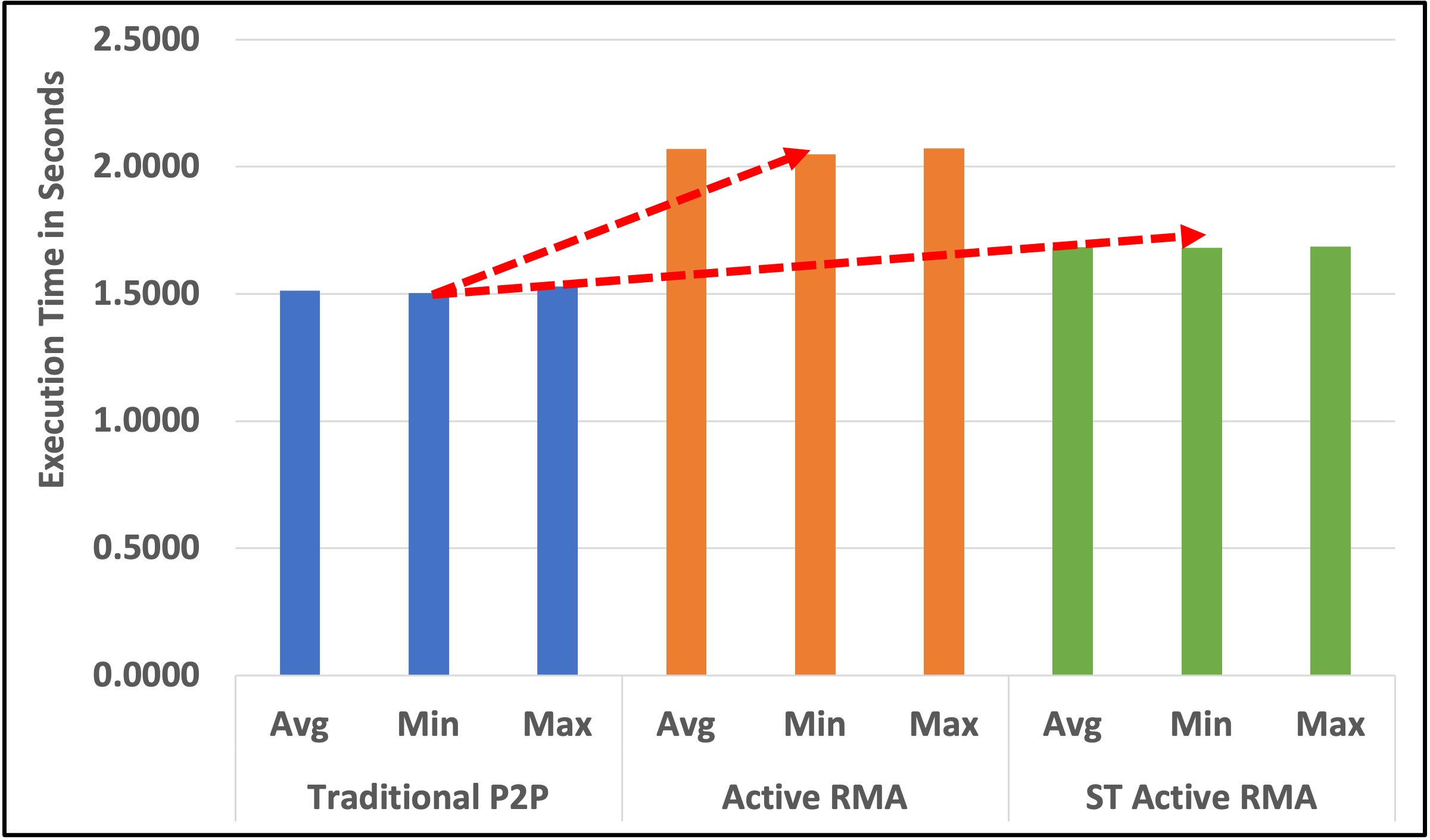}
  \caption{Multi-node comparison with traditional P2P.}
  \label{fig:inter-p2p}
\end{figure}

On the multi-node distribution, this test uses 64 MPI ranks distributed across
8 nodes, with 8 ranks per node. On each node, a one-to-one mapping
between MPI ranks and GCDs is enforced. Figure~\ref{fig:inter-p2p} shows the
results of the analysis.
In brief, the traditional P2P variant of the benchmark performs 27\% and 11\% better
than the active RMA and ST active RMA variants of the benchmark, respectively.
This performance behavior is significantly different from the comparisons shown in
Figure~\ref{fig:intra-p2p} within a single node.

The performance behavior shown in Figure~\ref{fig:inter-p2p} is not fully
understood, and analysis is in progress. Our initial assumption points to the
inherent performance characteristics of the triggered communication operations
and the usage of triggered put operations for inter-node signaling. Detailed
understanding on the performance characteristics of the triggered operations is
in progress, and usage of other forms of signaling is being considered.

\subsection{Overall Performance Inference}
Performance analysis from Section~\ref{subsec:over-perf} shows the potential
benefit of fully offloading the MPI control path into the GPU from the
application host process. Both multi-node (23\% improvement) and single-node
(36\% improvement) performance of ST active RMA is significantly better than the
active RMA version of the benchmark without GPU stream awareness.
Section~\ref{subsec:tops} and Section~\ref{subsec:merged-gpus} show the need for
using an effective throttling algorithm as well as merged GPU kernels in the ST
active RMA implementation.

While the relative comparison of the active RMA variant of the benchmark with
and without GPU stream awareness is used in most of the section, it is
established through Section~\ref{subsec:p2p-perf} that the ST RMA implementation
of the Faces benchmark shows either a significant improvement (single node) or an 11\%
performance hit (multiple nodes) versus the traditional P2P communication model.
Options for improving the multi-node performance of the ST active RMA variant
against the traditional P2P variant of the benchmark are being analyzed.

Overall, the major design goal of fully offloading the MPI control path into the
GPU is realized using the proposed ST active RMA communication scheme. The
performance results from analyzing the Faces microbenchmark show both the
benefits as well as the issues with the current implementation. 

The Comb~\cite{comb} communication microbenchmark was also ported to standard MPI active RMA and ST active RMA,
and the performance results were similar. For brevity those results are not included.

%% file: src/content/related.tex

\section{Other Proposal Considerations}\label{sec:other-considerations}
While the proposed ST solution uses MPI active RMA,
multiple other alternatives were considered.
\begin{enumerate}
    \item \textbf{Traditional P2P communication}
    \begin{itemize}
        \item As described by \citeauthor{namashivayam2022exploring}~\cite{namashivayam2022exploring},
        implementing MPI message matching semantics~\cite{hw_mm,bmm,keith-message-matching,Flajslik2016MitigatingMM,Bayatpour2016AdaptiveAD,Schonbein2018MeasuringMM,osti_hw_mm} for intra-node
        communication operations is not as simple as the design from
        Section~\ref{subsec:st-arma-intra-node}. A progress thread is required
        to emulate the deferred execution semantics for intra-node operations.
        \item Traditional P2P semantics do not enable aggregation, while aggregation
        is straightforward during RMA operation (\emph{MPIX\_Win\_complete\_stream}).
        API changes are required to support aggregation for P2P.
        \item As an initial design, a P2P-based ST solution was
        evaluated~\cite{namashivayam2022exploring}. For brevity, results of this effort are not included as
        part of this work. In brief, the ST with P2P solution showed limited performance
        improvements on use cases where only inter-node data transfers were
        involved, and only by using ST for just the send side. Using a progress thread to emulate the deferred execution
        semantics for use cases with intra-node transfers negated any
        potential performance benefits.
    \end{itemize}
    \item \textbf{Persistent P2P communication}~\cite{persistent-mpi,persistent-opt-mpi} allows aggregation, but it has
    similar implementation limitations for intra-node transfers as the
    traditional P2P interfaces.
    \item \textbf{Partitioned communication}~\cite{partioned-impl,grant2019finepoints} may have promise for pipelining
    communication to individual targets through triggering the operations from
    within a GPU kernel, but it provides no advantage for aggregating multiple messages to different neighbors.
    \item \textbf{One-sided Passive RMA communication}~\cite{mpi-rma} avoids tag
    matching and has potential benefits
    similar to the proposed solution. But the passive semantics make aggregation
    difficult, and the global fences make efficient synchronization difficult
    for nearest-neighbor communication use cases as discussed in
    Section~\ref{subsec:test-case}.
\end{enumerate}

\section{Related Work}\label{sec:related}
There is limited research performed on exploring options for introducing
GPU stream-awareness into message-passing programming models with deferred
execution semantics~\cite{tops-portal}. Most known works involve using a
progress thread for implementation~\cite{zhou2022mpix}. To the authors' knowledge,
this work describes the first
GPU-aware MPI implementation to offload all control and data operations
for both intra-node and inter-node communication
from the CPU to the GPU and NIC, allowing the CPU to proceed asynchronously
without any progress threads.

\citeauthor{offload-communication-control}~\cite{offload-communication-control} explores the basic building blocks
required for offloading communication control logic in GPU accelerated
applications into the GPU device. \cite{offload-communication-control} explores
hardware offload capabilities using NVIDIA GPUDirect Async~\cite{nvda-gdsync,nvda-gpudirect}
and InfiniBand Connect-IB network adapters~\cite{verbs-rdma}.

\citeauthor{gpu-verbs}~\cite{gpu-verbs} and \citeauthor{gpurdma}~\cite{gpurdma}
explore the possibilities of
GPUs handling the communication and control paths by hosting the Verbs
layer~\cite{verbs-rdma} in the GPU. Similarly, \citeauthor{nvshmem}~\cite{nvshmem} and
\citeauthor{rocshmem}~\cite{rocshmem} explore the options to host
communication
on the GPU using the OpenSHMEM programming model\cite{osm14}. This
research~\cite{gpu-verbs,gpurdma,nvshmem,rocshmem} is mostly done using the
NVIDIA Infiniband Host Channel Adapter.

%% file: src/content/conclusion.tex

\section{Conclusion}\label{sec:conclusion}
In this work, a new strategy uses \textit{stream-triggered
communication} to introduce GPU stream-awareness in MPI
without the need for CPU progress threads.
The proposed strategy extends MPI active target synchronization 
to allow an application to offload both the \textit{control} and
\textit{data} paths to the underlying implementation and hardware components,
and it avoids CPU-GPU synchronization at the GPU compute kernel
boundaries.

Description of the proposed implementation shows the ability to
fully offload the communication control path from the CPU to GPU. Performance
analysis using the Faces microbenchmark kernel shows 23-36\% improvement
in performance over the MPI active RMA communication model without GPU stream
awareness. Benefits in selecting an optimized throttling algorithm to manage
triggered operation resources and using merged GPU kernel strategies are also
discussed. Overall the performance results reflect the performance improvements
on applications with near-neighbor communication models like Faces for the
latency-bound regime of small messages.

While ST active RMA shows the expected performance improvements for
intra-node communication, inter-node performance does not show the
expected advantages over standard MPI P2P communication. 
Future work will investigate this performance shortfall
and endeavor to improve it.